# A compact analytical formalism for current transients in electrochemical systems

Pradeep R. Nair and Muhammad A. Alam

School of ECE, Purdue University, West Lafayette, IN **Abstract**: Micro and nanostructured electrodes form an integral part of a wide variety of electrochemical systems for biomolecule detection, batteries, solar cells, scanning electrochemical microscopy, etc. Given the complexity of the electrode structures, the Butler-Volmer formalism of redox reactions, and the diffusion transport of redox species, it is hardly surprising that only a few problems are amenable to closed form, compact analytical solutions. While numerical solutions are widely used, it is often difficult to integrate the insights gained to the design and optimization of electrochemical systems. In this article, we develop a comprehensive analytical formalism for current transients that not only anticipate the response of complex electrode structures to complicated voltammetry measurements, but also intuitively interpret diverse experiments such as redox detection of molecules at nanogap electrodes, scanning electrochemical microscopy, etc. The results from the analytical model, well supported through detailed numerical simulations and experimental data from literature, have broad implications for the design and optimization of nanostructured electrodes for healthcare and energy storage applications.

**I. Introduction**: Various forms of voltammetry (or chronoamperometry) constitute the basic characterization techniques for electrochemical systems and provide valuable information regarding the geometry and reaction constants of complex nanostructured electrodes. Since the landmark article by Nicholson and Shain in 1964 for numerical solutions of the voltammetry[1] problems, the field has witnessed tremendous research activity to unravel the dynamics of electrochemical processes at electrodes. As most problems are not amenable to closed form analytical solutions, various numerical simulation schemes had to be developed, instead, to analyze the response of electrochemical systems[2-4]. Nevertheless, analytic solutions often provide crucial and nontrivial insights regarding various sub-processes that help significantly in the design and further optimization. In this article, we develop a closed-form, compact analytic formulation for current transients at microelectrodes based on the well known reaction-diffusion formalism and use these solutions to study the dynamics of redox kinetics at microelectrodes. Apart from predicting the trends for classical electrode geometries like planar, cylindrical, and spherical surfaces, our analytical results also anticipate the important trends for fractal electrodes as well.

The article is arranged as follows: In Sec. II, we develop the detailed analytical formalism to study the current transients at generic electrodes. In Sec. III, we apply the analytical model to study (a) potential step voltammetry, (b) linear sweep voltammetry, (c) redox detection of molecules at nanogap electrodes[5, 6], and (d) scanning electrochemical microscopy[7, 8] (SECM). Although these topics are discussed/researched individually by using specialized numerical techniques and approaches[3], in general, the field lacks a common analytical framework that will help cross-connect the inferences and conclusions of various subtopics in a systematic way. For example, while closed-form analytical solutions are available for potential step voltammetry[3, 9-13], linear sweep voltammetry depends heavily on numerical simulations even for the simplest configurations. Here, the results are typically given in tabular format[1, 3] or require complex numerical integration[14]. Similarly, analytical description of redox detection of molecules is available only for simple geometries[5] while numerical simulations are required for other structures[15]. The scenario is no different for SECM, where the analytical expressions are limited to curve-fitting of numerical data[16, 17]. In this manuscript, we show that the above mentioned wide ranging electrochemical measurements can be uniquely and succinctly described through a comprehensive analytical formalism in terms of single mathematical concept of " Transient Diffusion Equivalent Capacitance (TDEC)". This mathematical concept of TDEC is a way of solving the diffusion equation and should not be confused with other physical capacitance such as double-layer capacitance, because after all it does not even have the same dimension! We validate our analytical model by comparing with numerical simulation and/or experimental data, as appropriate. Indeed, so long the basic assumptions of isotropic Fickian diffusion and heterogeneous redox reactions are valid, we are yet to find a system whose numerical solution is not anticipated by the analytical formula proposed in this manuscript. After discussing the implications and impact of the new model, we summarize the results in Sec. IV. Detailed derivations and numerical simulation methodology are reserved for appendices.

**(a)**

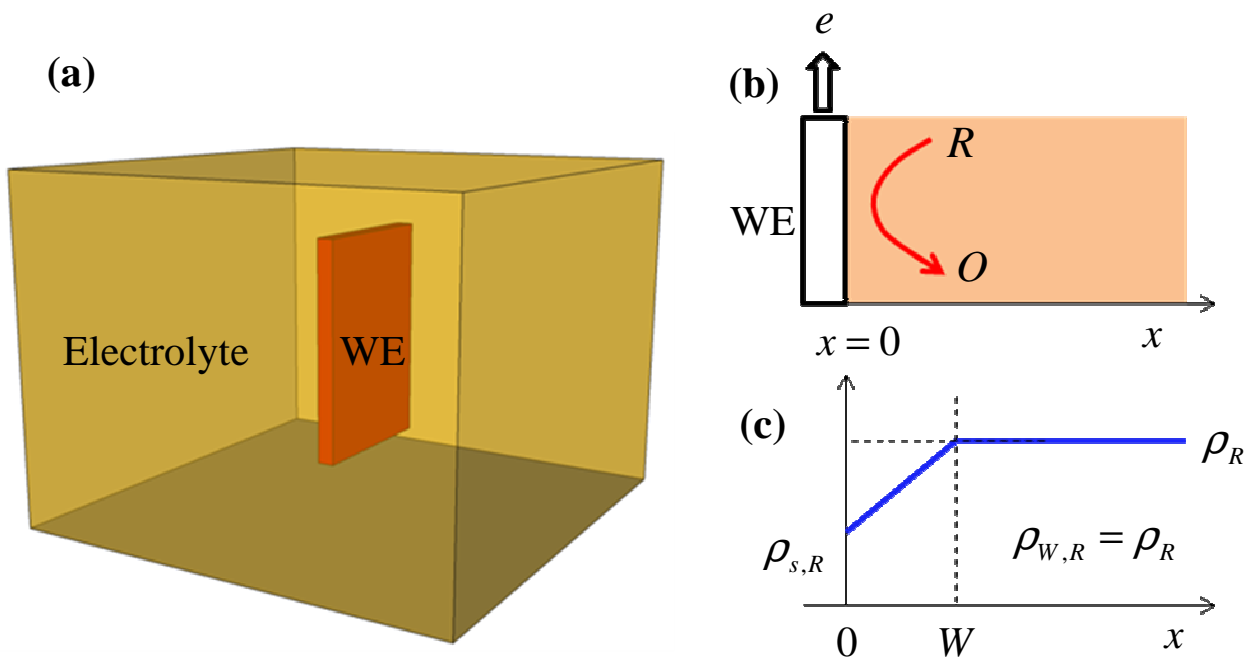

**Figure 1.** (a) Schematic of a WE introduced to an electrolyte. (b) The reaction $R \rightleftarrows O + e$ occurs at the electrode surface, with diffusion limited transport of R and O in the electrolyte and reaction rate at the electrode surface is dictated by the Butler-Volmer formalism. (c) Linear concentration profile assumed for the derivation of compact analytical model. Note that although the schematic and concentration profiles are shown for a planar electrode, the analytical formalism developed in this article is applicable for a wide variety of electrode structures.

## II. A. Model System

**II. A. Model System**: The model system consists of a working electrode (WE) immersed in a solution of target molecules (see Fig. 1a. For simplicity, a planar electrode is shown in the illustration, although the formalism we develop in subsequent sections will be applicable to a wide variety of nonplanar electrode structures). Oxidation/reduction of the redox species occur at the electrode surface, depending on the applied potential (against a reference electrode, usually Ag/AgCl). The electrochemical reaction at the electrode surface is characterized as

$$R \rightleftarrows O + e, \tag{1}$$

where R and O denote the reduced and oxidized species, respectively. The concentration of R near the electrode surface reduces as the reaction proceeds. Thus, the dynamics of electrochemical reaction can be described through a reaction-diffusion process which involves a redox reaction with net rate given by

$$R \rightleftarrows O; \qquad net\ rate: -k_F \rho_{s,R} + k_R \rho_{s,O}, \tag{2}$$

where $k_F$ and $k_R$ are the forward and backward reaction constants and $\rho$ denote the concentration of corresponding species. Note the convention for subscripts associated with $\rho$: the first subscript denotes the spatial location ('s' means WE *surface*), while the second denotes the reactant species. The rate constants, $k_F$ and $k_R$, depend on the potential of the working electrode and will be considered explicitly later (see eq. (13) , Sec. IIB). The concentration profiles for R and O are dictated by the diffusion equation

$$\frac{d\rho_{R,O}}{dt} = D_{R,O}\nabla^2\rho_{R,O},$$ (3)

where D is the diffusion coefficient of the corresponding species, and $\nabla^2$ denote the Laplacian operator in appropriate co-ordinate system. Equations (2)-(3) determine the dynamics of the system. As eq. (2) also represents the current density at the electrode surface, the net electrode current is given as

$$I = -q\int \left(rate\,of\,reaction\right)dS,$$ (4)

where the integration is over the electrode surface area, and q is the electronic charge. Assuming uniform distribution of reactants around the electrode and spatial homogeneity of reaction constants, Eq. (4) can be re-written as

$$I = A_e q\left(k_F\rho_{s,R} - k_R\rho_{s,O}\right)$$ (5)

where $A_e$ is the electrode surface area. Flux or mass conservation at the electrode surface leads to

$$D_R\nabla\rho_{s,R} = -D_O\nabla\rho_{s,O} = Eq.\left(2\right).$$ (6)

Equations (2)-(6) along with the initial conditions for $\rho_{R,O}$ determine the behavior of the model system described in this section.

Note that the model system of eqs. (2)-(6) ignore the transient effects due to electrolyte double layer charging process and/or any uncompensated resistance. The influence of such effects on current transients is elaborately discussed in literature[3]. For redox species, we assume diffusion limited transport and neglect convection and migration (i.e., migration of ions in an electric field). Usually, this assumption is valid as an excess electrolyte is almost always present at much higher concentrations in most systems. Any associated electric field effects are strongly screened by the electrolyte (the debye screening length is of the order of a few nm for mM electrolyte concentrations[18]). Moreover, our numerical simulations (results not discussed in this article) with an additional migration component (proportional to the electric field) in the presence of excess electrolyte show negligible deviations from the analytical solutions. Indeed, recent literature also indicates that results based on diffusion analysis can be applied to nanoscale electrodes in the presence of excess electrolyte[19] (although there has been some reports of non-linear transport phenomena near nanoscale electrodes[20]). Therefore we believe that while the closed form analytical solutions discussed in this article might have to be refined to reflect the complexities associated with electrolyte screening, finite size of molecules, anisotropy of diffusion/reaction, etc., the corrections are likely to be relatively minor and the solutions provided here can be interpreted as a good approximation to the full solution.

**II. B. Analytical solution:** Here, we first develop an analytic solution for the transient current at a planar WE and then extend it to various other electrode geometries. Consider a 1-D planar system (semi-infinite) with WE introduced at t=0 (Fig. 1b, with cross-sectional area $A_{planar}$) in the presence of an analyte solution with only R species at a concentration $\rho_R$ (O molecules are absent at t=0, see Fig. 1c). As the reaction proceeds, R is converted to O at the electrode surface and diffusion dictates the transport of molecules in the system. After time 't', let us assume that the component R has been depleted to a distance of W (the *diffusion layer thickness*, see Fig. 1c.), so that bulk concentration is maintained for $x > W$. Finite reaction rate at WE and the diffusion process determine the non-zero concentration of R at $x = 0$. We assume linearly varying concentration profiles for R, $0 < x < W$, where W is the diffusion layer thickness (For planar systems, the concentration profiles are not exactly linear, but vary as $erf\left(x/\sqrt{Dt}\right)$, ref.[3]. Note that we make this assumption to derive a closed form analytical solution for current transients, which are then compared with detailed numerical simulations and literature in Sec. III).

The integrated current until time 't' is given by the net amount of species R that was oxidized (i.e., the depleted triangle in Fig. 1c, also see discussion associated with Eq. 1.4.30, Sec. 1.4.3 of ref.[3], the textbook by Bard and Faulkne**r**), so that

$$\int_0^t I d\tau = q A_{planar} \frac{W}{2} \left(\rho_{W,R} - \rho_{s,R}\right), \tag{7}$$

which on differentiation with time 't' implies,

$$\frac{I(t)}{q A_{planar}} = \left(\rho_{W,R} - \rho_{s,R}\right) \frac{dW}{2dt} + \frac{W}{2}\left(\frac{d\rho_{W,R}}{dt} - \frac{d\rho_{s,R}}{dt}\right). \tag{8}$$

The second term on the right hand side of eq. (8) can be evaluated by assuming linear diffusion (see Fig. 1c), i.e.

$$\frac{d\rho_{W,R}}{dt} = \frac{d\rho_{W,R}}{dW}\frac{dW}{dt} \approx \frac{\left(\rho_{W,R} - \rho_{s,R}\right)}{W}\frac{dW}{dt}. \tag{9}$$

| Electrode | $A_e$ | $C_{D,SS}$ | $C_{D(t)}$ |
|---|---|---|---|
| Planar | $A_{planar}$ | $\dfrac{A_e D}{W}$ | $\dfrac{A_e D}{\sqrt{2Dt}}$ |
| Cylindrical | $2\pi a_0$ | $\dfrac{2\pi D}{\log\left(\dfrac{W+a_0}{a_0}\right)}$ | $\dfrac{2\pi D}{\log\left(\dfrac{\sqrt{4Dt}+a_0}{a_0}\right)}$ |
| Spherical | $4\pi a_0^2$ | $\dfrac{4\pi D}{a_0^{-1}-\left(W+a_0\right)^{-1}}$ | $\dfrac{4\pi D}{a_0^{-1}-\left(\sqrt{6Dt}+a_0\right)^{-1}}$ |
| Micro disk | $\pi a_0^2$ | $4Da_0+\dfrac{\pi D a_0^2}{W}$ | $4Da_0+\dfrac{\pi D a_0^2}{\sqrt{2Dt}}$ |
| Array | $2\pi a_0$ | $\dfrac{2\pi D}{\log\left(\dfrac{\sinh\left(2\pi\gamma\left(W+a_0\right)\right)}{\pi a_0\gamma}\right)}$ | $\dfrac{2\pi D}{\log\left(\dfrac{\sinh\left(2\pi\gamma\left(\sqrt{2Dt}+a_0\right)\right)}{\pi a_0\gamma}\right)}$ |
| Parallel NW | $2\pi a_0$ | $\dfrac{\pi D}{\log\left(\dfrac{W}{a_0}+\sqrt{\left(\dfrac{W}{a_0}\right)^2-1}\right)}$ | |

**Table 1.** Diffusion equivalent capacitance for various electrode geometries. W denotes the electrode spacing and $a_0$ is the radius of corresponding electrode.

As the redox reaction at the electrode proceeds, the species R will be depleted further away from the electrode surface. Now the *diffusion layer thickness* is given as $W \sim \sqrt{2D_R t}$ by classical diffusion theory[21] (note that various other definitions are routinely used for this parameter, see Sec. 5.2.1 of ref.[3]), $dW/dt = D_R/W$. Using this relation and eq. (9), we can rewrite eq. (8) as

$$\frac{I(t)}{qA_{planar}} = \frac{D_R}{W}\left(\rho_{W,R}-\rho_{s,R}\right)-\frac{W}{2}\frac{d\rho_{s,R}}{dt}. \tag{10}$$

Equation (10) represents the current at a planar WE at time 't' for the concentration profile shown in Fig. 1. Note that the seeds of the above analysis are already present in the classic textbook of Bard and Faulkner (Sec. 1.4.3, ref.[3]) as a pedagogical tool for planar systems. However, here we generalize the idea to a much broader context of complex system with arbitrary configuration of electrode geometries, as shown in the following discussions.

Using the terminology provided in Table 1 with $\rho_{W,R} = \rho_R$, eq. (10) can be re-written for electrodes of arbitrary geometry as

$$I(t) = qC_{D,SS}\left(\rho_R - \rho_{s,R}\right) - \frac{qA_e W}{2}\frac{d\rho_{s,R}}{dt},\tag{11}$$

where $A_e$ denotes the electrode area. For example, $A_e = A_{planar}$ and $C_{D,SS} = \dfrac{A_{planar} D_R}{W}$ for planar systems, but the values will be different for other electrode geometries. The parameter $C_{D,SS}$ is known as diffusion equivalent capacitance[21]. Note that $C_{D,SS}$ has the same functional form as the electrical capacitance between two electrodes separated by a distance W, except that the dielectric permittivity is replaced by the diffusion coefficient, D (see Table 1). This concept of diffusion equivalent capacitance follows from the elegant analogy of diffusion problems in bio-chemical literature to Laplace's equation of electrostatics and it provides rich insights and powerful analysis techniques that have been used by biophysics community with significant success[22] (see appendix A for a detailed discussion on $C_{D,SS}$). The introduction of $C_{D,SS}$ is an important step as it allows us to generalize eq. (10) beyond 1D planar systems (note that we transition from presumed linear concentration profiles for deriving eq. (10)), to eq. (11) which holds good for electrodes with arbitrary geometry[23] (e.g., cylindrical, spherical electrodes, see the corresponding $C_{D,SS}$ in Table 1 and appendix A).

Equation (11) represents the transient current at a micro electrode with a surface concentration $\rho_{s,R}$ and bulk concentration $\rho_R$ at a distance W from the electrode. However, both $\rho_{s,R}$ and W are time dependent. $\rho_{s,R}$ is dictated by the rate of redox reaction (eq. (2)) and diffusion process (eq. (3)), while W varies as $\sqrt{t}$, as discussed earlier. The time dependent variation of $\rho_{s,R}$ can be accounted through a generalized perturbation approach (using the eqs. (6) and (11), assuming $D_R = D_O = D$ without loss of generality, see Appendix B for a detailed derivation), and we find that the transient current at electrodes of arbitrary shape is given by

$$I(t) = qA_e\rho_R\left(\frac{k_F}{1 + \dfrac{A_e}{C_{D(t)}}\left(k_F + k_R\right)}\right) - qA_e\frac{W}{2}\frac{d\rho_{s,R}}{dt}.\tag{12}$$

Note that eq. (12) explicitly incorporates the effects of finite reaction rates $k_F$ and $k_R$ (see eq. (2)) The major conceptual addition to eq. (11) is the introduction of $C_{D(t)}$, the time dependent version of diffusion equivalent capacitance $C_{D,SS}$ (TDEC, see Appendix B for a detailed discussion). As discussed before, $C_{D,SS}$ (and hence $C_{D(t)}$) is obtained by exploiting the analogy between diffusion problem and electrostatic systems. We note that the functional form of $C_{D(t)}$ is exactly the same as the electrostatic capacitance of the electrode system with two essential changes – (i) dielectric permittivity is replaced with diffusion coefficient D, (ii) the spatial separation parameter W is replaced by $\sqrt{2nDt}$, the diffusion distance (n is an integer)[23]. For example, consider planar systems (see Table 1). It is well known that the electrostatic capacitance[24] of a system of planar electrodes is given as $A_e\varepsilon / W$, where $\varepsilon$ is the dielectric permittivity. Now as suggested before, by replacing $\varepsilon$ with D, and W with $\sqrt{2Dt}$, $C_{D(t)}$ of planar systems is given as $A_e D / \sqrt{2Dt}$. $C_{D(t)}$ for other systems described in Table 1 directly follows this methodology. We emphasize that as TDEC is a mathematical technique to solve transient diffusion equation by exploiting its analogy to electrostatic problem, $C_{D(t)}$ has appropriate dimension for the problem being solved (not that of a typical electrical capacitance).

The constants $k_F$ and $k_R$ are bias dependent and in the Butler-Volmer formalism[3] are given as

$$k_F = k_0 e^{(1-\alpha)f(E_A - E_0)}$$
$$k_R = k_0 e^{-\alpha f(E_A - E_0)}$$

(13)

where $k_0$ is the heterogenous rate constant, $\alpha$ is the transfer coefficient (usually, $\alpha = 0.5$), $f = F/RT$, F is the Faraday's constant, R is the universal gas constant, T is the temperature, $E_A$ is the applied bias at the WE and $E_0$ is the formal potential of the reaction. We assume that any effects of electric field on the reaction rates are factored in through appropriate parameterization[25]. Using eq.(13), the general solution for transient current at a WE is given as

$$I(t) = qA_e\rho_R \left( \frac{e^{(1-\alpha)f(E_A - E_0)}}{\dfrac{1}{k_0} + \dfrac{A_e}{C_{D(t)}}\left(e^{(1-\alpha)f(E_A - E_0)} + e^{-\alpha f(E_A - E_0)}\right)} \right) - qA_e\frac{W}{2}\frac{d\rho_{s,R}}{dt}.$$

(14)

We assume that the bias applied to the WE is of the form $E_A = E_i + vt$, where $E_i$ is the starting potential and $v$ is the sweep rate. The rate of change of surface concentration, $d\rho_{s,W}/dt$ is influenced by the applied bias $E_A$ and the sweep rate $v$. In general, we find that

$$\frac{d\rho_{s,R}}{dt} = -\rho_R \frac{v\beta f e^{\beta f(E_A - E_0)}}{\left(1 + e^{\beta f(E_A - E_0)}\right)^2} \qquad (15)$$

is a good approximation (see appendix C for derivation, $\beta$ is a factor to capture the deviation from Nernst behavior due to the diffusion limited transport of molecules, $0 < \beta < 1$). Similarly, to evaluate W in the second term on the RHS of eq. (14), we realize that for $E_A < E_0$, redox reaction at WE is negligible and $W \sim 0$, on the other hand for $E_A >> E_0$, W approaches $\sqrt{2Dt}$. This asymptotic behavior can be captured by using an interpolation function of the form $W \sim \sqrt{2Dt} e^{\beta f(E_A - E_0)} \left(1 + e^{\beta f(E_A - E_0)}\right)^{-1}$. Using eqs. (14) and (15), we find that the transient current for a time varying potential at the electrode is given as

$$I(t) = qA_e\rho_R \left( \frac{e^{(1-\alpha)f(E_A - E_0)}}{\frac{1}{k_0} + \frac{A_e}{C_{D(t)}}\left(e^{(1-\alpha)f(E_A - E_0)} + e^{-\alpha f(E_A - E_0)}\right)} \right) + qA_e\rho_R \frac{\sqrt{2Dt}}{2} \frac{v\beta f e^{2\beta f(E_A - E_0)}}{\left(1 + e^{\beta f(E_A - E_0)}\right)^3}. \qquad (16)$$

Equation (16) is the key result of this article and provides closed form analytical solutions for the transient response of electrodes with arbitrary geometry. Note that the geometry of the electrode and the transient diffusion of reactants are incorporated through the parameter $C_{D(t)}$. Appropriate $C_{D(t)}$ (see Table 1) and eq. (16) allow us to study the dynamics of various electrochemical processes at nanostructured electrodes. In general, eq. (16) implies that the transient current at WE consists of two components: (i) A component due to the bulk diffusion effects (first term on the RHS, called *the diffusion component*), and (ii) second component due to the rate of consumption of reduced species R at the WE (second term on the RHS, called as *the reaction component*). The relative magnitude of these two components has interesting implications on the transient behavior of micro and nanostructured electrodes, as discussed in section III.A.

The above analysis completes the derivation for transient currents at microelectrodes. Note that we have, so far, not placed any restriction on the shape of the electrode. Moreover, the analysis has relied only on simple, physical arguments that obviate the need for complex Laplace-Transform approach, series solution, or complicated numerical simulations, typically found in electrochemical textbooks[3]. As mentioned before, an approach similar to our analysis is given as a semi-empirical proof for Cottrell equation in ref.[3] (i.e., restricted only for potentiostatic measurements of planar systems). However, our approach, based on transient diffusion equivalent capacitance, has been generalized to other complex systems with a wide variety of electrode shapes and operating conditions. The conceptual power of our approach lies in the fact that (i) it can integrate multiple topics, that may otherwise appear only weakly related, into a single conceptual framework, and that (ii) it transforms a complex, transient, unsteady state mass transport problem into a benign (and often already solved) problem of steady-state transport.

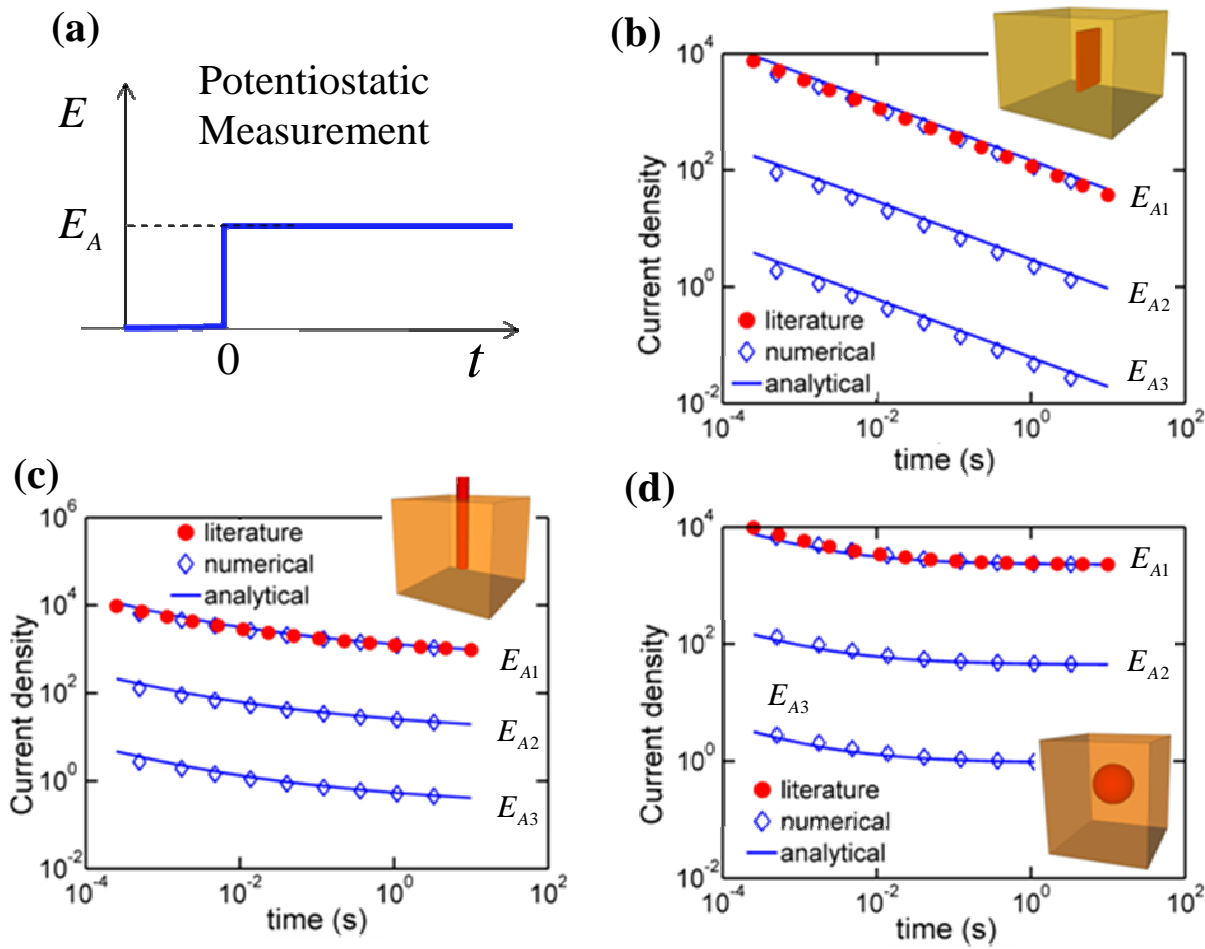

**Figure 2.** Comparison of the analytic model with results from literature for potentiostatic measurement. (a) Schematic of a potentiostatic measurement. At $t = 0$, a voltage step is applied at the electrode. Transient response of (b) planar, (c) cylindrical $(a_0 = 1 \mu m)$, and (d) spherical electrode $(a_0 = 2 \mu m)$. The solid symbols represent results from literature (refs.[9-11]), open symbols represent numerical simulation results of this work ( $E_{A1} = E_0 + 0.2V$;  $E_{A2} = E_0 - 0.1V$; $E_{A3} = E_0 - 0.2V$, see Appendix D for details), and solid lines represent analytical results (eq. 17). Note that apart from reproducing the results form the literature for ideal electrodes with large bias ($E_A > E_0$), the analytical results also match the numerical simulation results for small bias ($E_A < E_0$).

**III. Results:** We now illustrate the validity and generality of our approach by comparing it against a set of well known electrochemical experiments. We first consider (a) potential step voltammetry and (b) linear sweep voltammetry. In the later sections we illustrate the generality of eq. (16) by considering complex scenarios like (c) redox detection of biomolecules using multiple electrodes and (d) scanning electrochemical microscopy. This heuristic validation does not exclude the possibility that there may be other problems that the analytical solutions cannot fully encapsulate, but the solution will nonetheless offer significant insights into the problem. We start with the potential step voltammetry:

### III. A. Potential Step Voltammetry (PSV):

In PSV, a step-bias $E_A$ is applied to the WE at time t=0 (see Fig. 2a). Any potential step bias, in general, can be visualized to consist of two segments: (a) ramp process in which the electrode potential is increased to $E_A$ such that $0 < t < 0^+$, $v \to \infty$ ($0^+$ denotes the time required by the system to raise the electrode potential to $E_A$) and (b) the potential is held fixed at $E_A$ for the rest of the measurement such that $t > 0^+$, $v = 0$. The reaction component (second term on the RHS of eq. (16)) is negligible during part (a) as $t \sim 0$. The same component is negligible during part (b) as $v = 0$. Hence the transient current for PSV (using eq. (16)) is given by

$$I(t) = qA_e\rho_R \left( \frac{e^{(1-\alpha)f(E_A-E_0)}}{\dfrac{1}{k_0} + \dfrac{A_e}{C_{D(t)}}\left(e^{(1-\alpha)f(E_A-E_0)} + e^{-\alpha f(E_A-E_0)}\right)} \right). \tag{17}$$

Analytical results are available in literature for the response of planar, cylindrical, and spherical electrodes for infinite reaction rates at WE (i.e., $k_0 \to \infty$, $E_A \gg E_0$). Figure 2b-d shows that eq. (17) compares well with literature (solid symbols, ref.[3, 9-11]) and numerical simulation results (open symbols, refer Appendix C for details) for $E_A \gg E_0$. At the same time, eq. (17) accurately predict the response for $E_A \ll E_0$ as well (biases $E_{A2}$ and $E_{A3}$ in Fig. 2), thus providing a general formalism to study the transient response of nanostructured electrodes.

Equation (17), with appropriate $C_{D(t)}$, anticipates well-established results from traditional literature on current transients at microelectrodes. The diffusion limited current at an electrode (with $E_A \gg E_0$ in eq. (17)) is given as

$$I(t) \approx q\rho_R C_{D(t)}. \tag{18}$$

For planar systems (refer Table 1) with $C_{D(t)} = \dfrac{A_{planar}D}{\sqrt{2Dt}}$ eq. (18) indicates that $I(t) \propto t^{-\frac{1}{2}}$, the famous Cottrel current[3, 9] (Fig. 2b). Note that since we use $\sqrt{2Dt}$ as the definition for diffusion layer thickness, eq. (18) for planar systems differ from Cottrell expression by a factor of $\sqrt{\frac{2}{\pi}}$. However, as discussed earlier, this definition is not unique (ref.[3]). The same formalism with appropriate $C_{D(t)}$ predicts the $\sim \dfrac{1}{\log(t)}$ response of cylindrical electrodes[10] (Fig. 2c). For spherical systems at large 't', $C_{D(t)} = \dfrac{4\pi D}{a_0^{-1} - \left(\sqrt{6Dt} + a_0\right)^{-1}} \sim 4\pi D a_0$. Eq. (18) now reduces to

$I(t \to \infty) = q\rho_R 4\pi D a_0$, the well established diffusion limited current towards a spherical electrode[11] (Fig. 2d). Similarly, eq. (18) with appropriate $C_{D(t)}$ compares well with the empirical formula in literature for disk microelectrodes[12]. For such electrodes, eq. (18) predicts that the diffusion limited current (i.e., as $t \to \infty$, using $C_{D(t)}$ from Table 1) is $I(t \to \infty) = q\rho_R 4 D a_0$, in accordance with literature[13]. It is well known that the current density at a planar electrode is much lower compared to spherical and cylindrical electrodes due to the diffusion limited transport of reactants[3]. In addition to predicting the results for diffusion dominated regime, eq. (17) accurately anticipates the results for the reaction dominated regime. For very small $k_0$, eq. (17) reduces to $I(k_0 \to 0) = q\rho_R A_e k_0 e^{(1-\alpha)f(E_A - E_0)}$, i.e., the reaction current is proportional to electrode area and varies exponentially with the applied bias.

Equation (17) provides a unique, compact, and generic formalism to understand and predict the current transients on a wide variety of microelectrodes. Note that the effect of finite reaction rates are incorporated in this formalism and can be extended to electrodes of arbitrary shape with appropriate parameterization of diffusion equivalent capacitance, $C_{D(t)}$. Although electrodes with size in $\mu m$ regime were used for validation purposes in Fig. 2, the model is equally valid for nanoscale electrodes (so long the presumption of Fickian diffusion is valid). Further, the model accurately anticipates electrochemical response at Nernst equilibrium, as and when it occurs as a function of various parameters like $k_0$, $E_0$, and $D$ (e.g., biases $E_{A2}$ and $E_{A3}$ in Fig. 2). We will now discuss the applicability of the new model to a more complex experiment, i.e., linear sweep voltammetry.

**III (B). Linear Sweep Voltammetry (LSV):** In LSV, the bias applied to the WE is a function of time, i.e., $E_A = E_i + vt$, where $E_i$ is the starting potential and $v$ is the sweep rate (we assume $E_i = 0$ for convenience, see Fig. 3a). The transient current in this case is given by eq. (16). Figure 3b-d shows that, once again, the closed form analytical result, eq. (16), reproduces the results from literature as well as numerical simulation (see Appendix D for details) results very well for planar, cylindrical and spherical electrodes. Literature data for planar and spherical electrodes are from ref.[3] (numerical results provided in a tabulated format) while for cylindrical electrode is from ref.[14] (that require complex numerical integration). Note that the planar electrodes clearly show a peak in transient current while such peaks are not so prominent in cylindrical and spherical systems[3]. Although analytical solutions in the form of series summation, semi-integrals, etc., are well known specific cases[26-29], eq. (16) provides a general solution protocol with compact analytical solutions for PSV/LSV of arbitrary electrode configurations and the model developed in this article allows experimentalists to anticipate their experimental trends without resorting to numerical simulations. Our results also predict all qualitative features of transient current as discussed below.

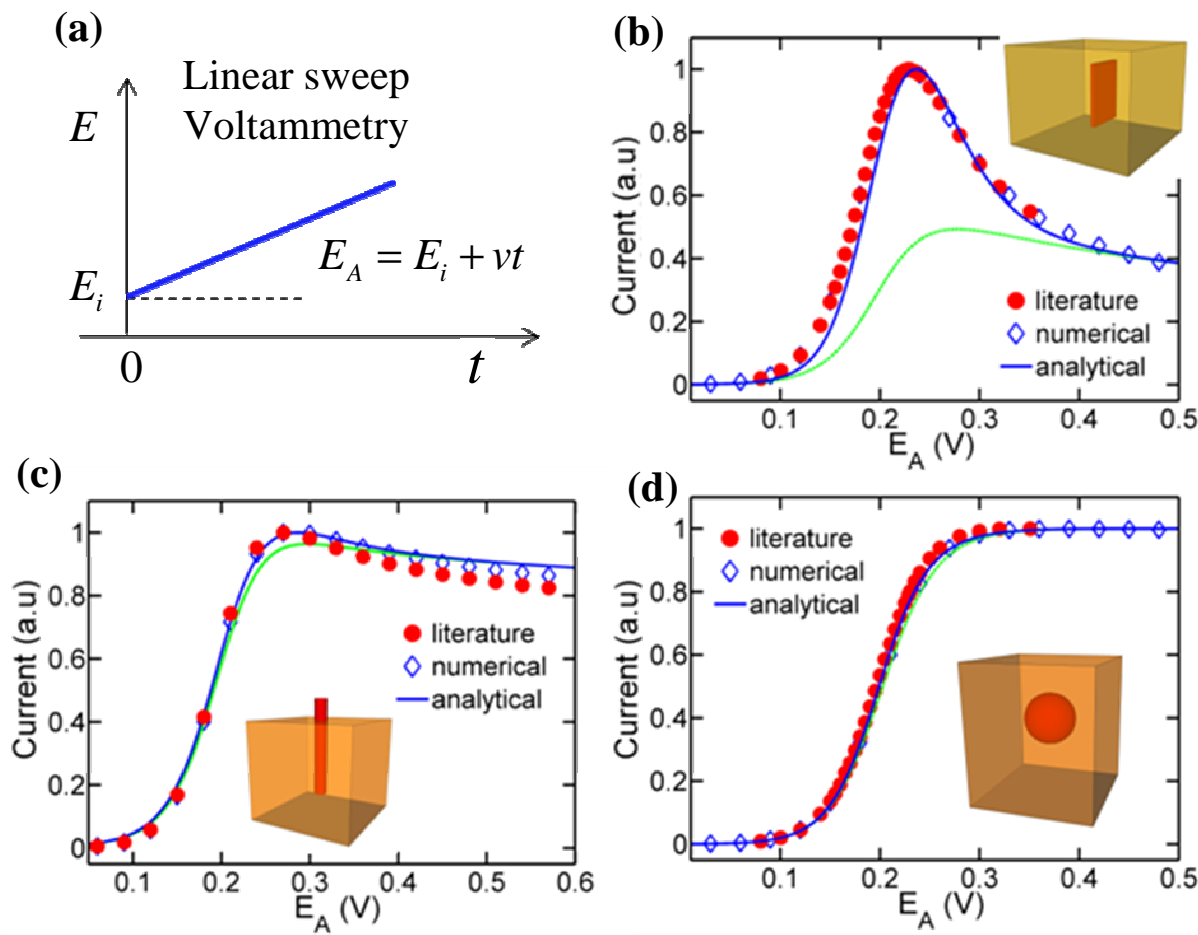

**Figure 3.** Comparison of the model with the results from literature for LSV measurement. (a) Schematic of a LSV measurement. At $t = 0$, a linearly increasing voltage is applied at the electrode $\left( v = 60\,mVs^{-1} \right)$. Comparison of results for (b) planar, (c) cylindrical $\left( a_0 = 1\mu m \right)$, and (d) spherical electrode $\left( a_0 = 2\mu m \right)$. The solid symbols represent results from literature (ref. [3,14], open symbols represent numerical simulation results of this work and lines represent analytical results. The solid (blue) curves indicate the total current, eq. (16), while the grey (green) lines represent the diffusion component (first term on the RHS of eq. (16)). Note that the reaction component is of comparable magnitude to the diffusion component only for planar system. For cylindrical and spherical systems, the diffusion current density is so high that it dominates the total response. For comparison purposes, the parameters are so chosen that $k_0 \left( Df v \right)^{-0.5} > 10$ (ref. [3], see Appendix C for simulation details).

**Dimensional effects on LSV**: It is a well known that spherical electrodes display quasi-steady state behavior during LSV (Fig. 3c), while planar electrodes exhibit well defined peaks with significant transient overshoot[3] (Fig. 3b). Although, these effects are usually attributed to the dimensional effects of diffusion towards microelectrodes, eq. (16) provides interesting insights to

Nair and Alam, Purdue University (2011)

this curious phenomenon. For planar systems, reaction current density is of comparable magnitude to the diffusion current component (see Fig. 3b. The grey (green) curve indicates the diffusion component and the solid (blue) curve denotes the total current). Therefore, as the reaction component is significant only at $E_A \sim E_0$, there is a well defined peak at the corresponding bias condition for the planar systems. However, since the diffusion flux in spherical electrode is orders of magnitude larger than the reaction component, spherical electrodes exhibit quasi-steady state behavior during LSV. Only for extremely fast voltage sweeps, can the reaction component for spherical electrode become comparable to its diffusion component, and lead to peaks in transient current comparable to those in planar electrodes[3].

**Dependence of LSV on sweep rate, $v$** : Eq. (16) can be expressed in terms of $E_A$, $\left( E_A \left( t \right) = vt \right)$ as

$$I(t) \approx q\rho_R C_{D(t)} \frac{1}{1+e^{-f(E_A-E_0)}} + qA_e\rho_R (Dv)^{1/2} \frac{\sqrt{2E_A(t)}}{2} \frac{\beta fe^{2\beta f(E_A-E_0)}}{\left(1+e^{\beta f(E_A-E_0)}\right)^3}. \qquad (19)$$

As mentioned before, the first term on the RHS represent the diffusion component while the second term represents the reaction component. Note that the second term (reaction component) is always proportional to $v^{1/2}$ while the sweep rate dependence of diffusion component (second term) depends on electrode geometry. For planar systems, using Table 1, as $C_{D(t)} = D(2Dt)^{-0.5} = (Dv)^{0.5} \left( E_A(t)/2 \right)^{-0.5}$, the diffusion component (first term) is proportional to $v^{1/2}$. Hence, eq. (19) indicates that for planar systems, $I(t) \propto \sqrt{Dv}$, in accordance with well established results of LSV[3]. For spherical systems, we find that $C_{D(t)} \sim t^0 \sim (E/v)^0$. As described in the previous section, the diffusion component is much larger than the reaction component and hence LSV is independent of $v$ for spherical systems[3]. Again, at extremely fast sweeps, spherical systems would also exhibit $v^{0.5}$ behavior as they resemble planar systems at short time 't'. The same analysis also predicts that while $I(t) \propto \sqrt{D}$ for planar systems, $I(t) \propto D$ for spherical systems, again reproducing well known trends in literature[3].

**Microelectrode arrays**: Periodic arrays of microelectrodes (Fig. 4a) are often used for sensitive calibration of LSV, yet the problem has only been addressed by numerical solution[30]. Instead, the analytical formalism can be used to describe the qualitative features of LSV with microelectrode arrays. The important parameters that characterize a microelectrode array are the size of the individual electrodes and the spacing between them. For such an array of cylindrical electrodes the appropriate expression for $C_{D(t)}$ is given as[24]

$$C_{D(t)} = \frac{2\pi D}{\log\left(\sinh\left(2\pi\left(\sqrt{2Dt}+a_0\right)\gamma\right)/\pi a_0\gamma\right)} \qquad (20)$$

where $\gamma$ denotes the density of NW array. For a given sweep rate, the spacing of the electrodes or rather the density dictates the LSV characteristics. Accordingly, as shown in Fig. 4c-d, at high densities, the array shows a pronounced peak similar to that of a planar system (Fig. 4c), while at low densities, the array response is analogous to that of a cylindrical system (Fig. 4d). Note that the widely varying behaviors of microelectrode arrays are accurately predicted by our approach and this formalism could provide an alternate methodology to the numerical simulation based diffusion domain analysis[30] for nanostructured electrodes.

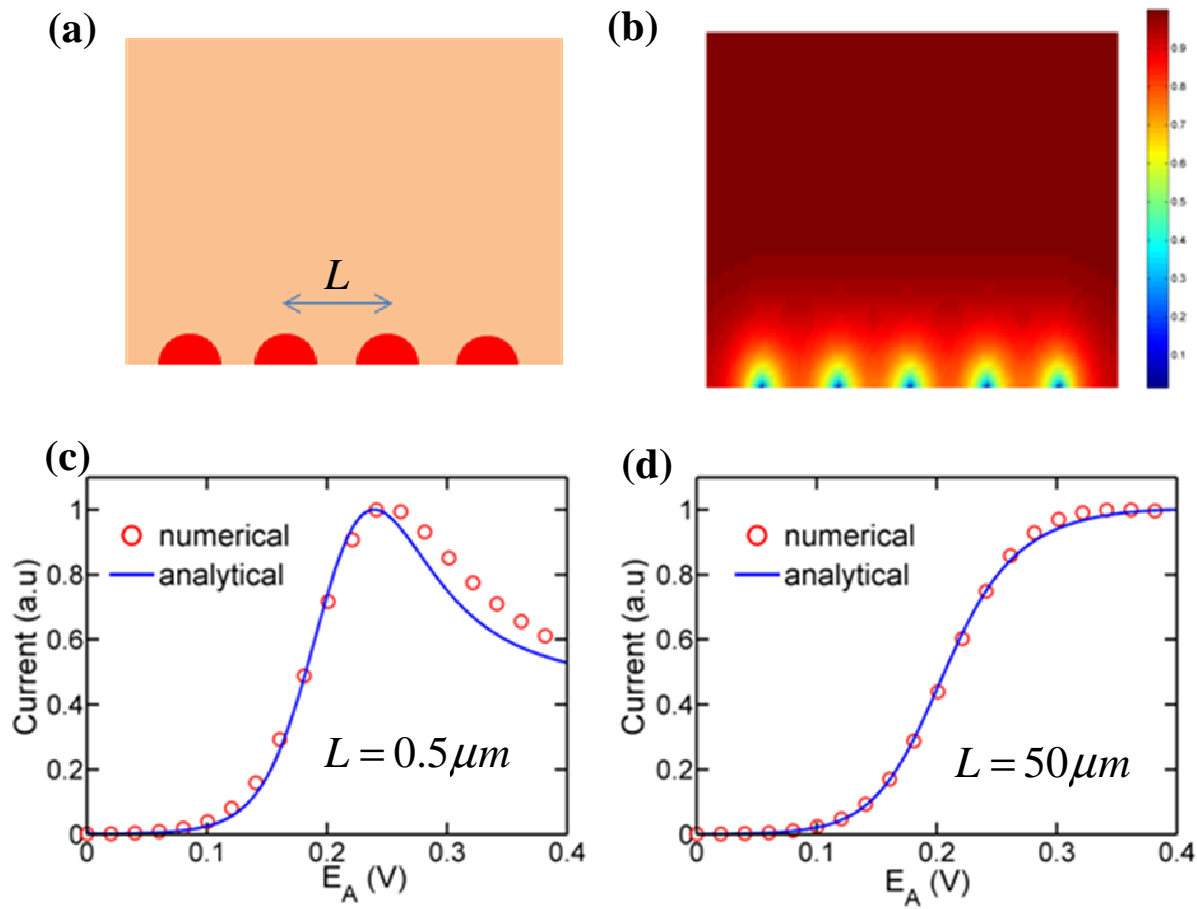

**Figure 4.** Application of the model to nanoelectrode array. (a) 2D Cross-section of an array where L denotes spacing between individual cylindrical electrodes ($a_0 = 20nm$) (b) Concentration profile of R during LSV. Note that the individual electrodes are shown in blue. (c) and (d) indicate the response of two arrays with different electrode spacing. Note that the same analytical formula predicts widely different behavior of two systems.

**Fractal electrodes**: Finally, we explore the problem of fractal electrode through the analytical formulation developed in the previous sections (see Fig. 5a). The response of fractal electrodes to voltammetry has been a topic of active research for several decades. In fractal electrodes, the electrode spacing is according to a power-law distribution dictated by the fractal dimension $D_F$.

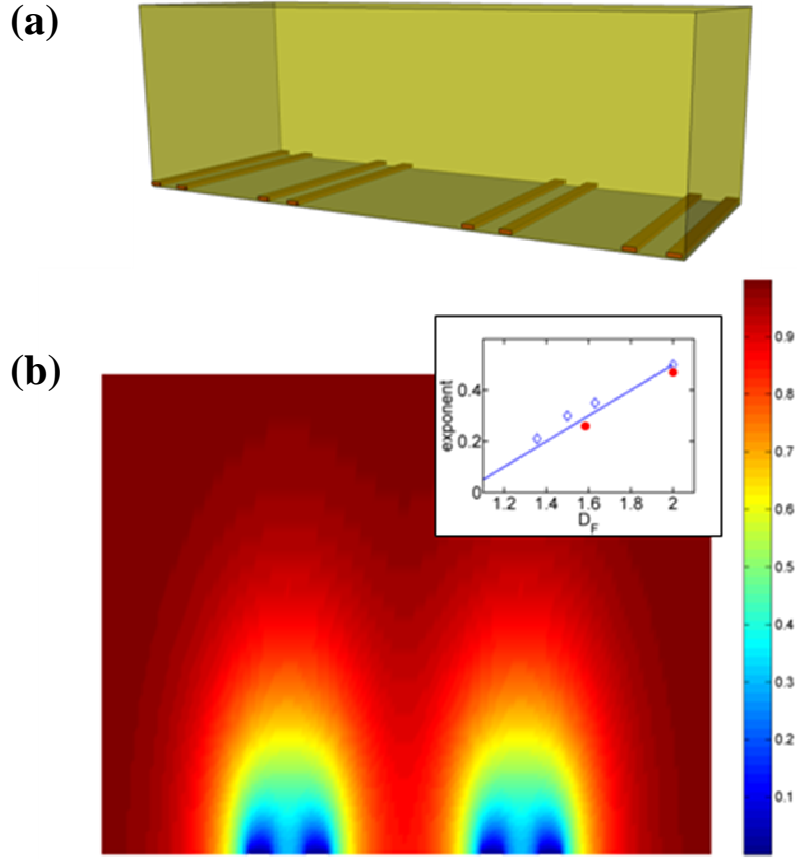

**Figure 5.** Model predictions for fractal electrodes. (a) Schematic of a fractal array which consist of thin rectangular strips. (b) Concentration profile of R during LSV for a structure with $D_F$=1.63. Inset shows comparison of sweep rate scaling exponents: experimental data from literature (ref.[30], solid symbols), numerical simulation results of this work (open symbols), and analytical results ($\frac{D_F - 1}{2}$, solid line).

This unique spatial distribution of fractal electrodes is also reflected in their current transients (as a power law in time). Accordingly, two remarkable and well known results from current transients at fractal electrodes are[31-33] : In a PSV, $I(t) \propto t^{\frac{1-D_F}{2}}$, and in LSV, the current scales with sweep rate as $I \propto v^{\frac{D_F-1}{2}}$. Both these well known trends[31-33] are reproduced in our analytic framework. For fractal systems $(1 < D_F < 3)$, we have $C_{D(t)} \propto t^{\frac{1-D_F}{2}}$ (or equivalently, $C_{D(t)} \propto \left(E/v\right)^{\frac{1-D_F}{2}}$ ref.[34, 35]). Using the eqs. (18) and (19), we find that $I(t) \propto t^{\frac{1-D_F}{2}}$ for PSV and $I(t) \propto v^{\frac{D_F-1}{2}}$ for LSV, respectively, thus reproducing well known results on current transients towards fractal electrodes. Figure 5b shows the concentration profile near a fractal electrode, while the inset shows a comparison of sweep rate scaling exponents from numerical simulations with experimental results from the literature[33]. Both the simulations and experiments follow the trends predicted by analytical results. Note that our model, through a general analytical framework, predicts both the transient and steady state LSV of a wide variety of electrodes, a significant improvement on previous literature[36].

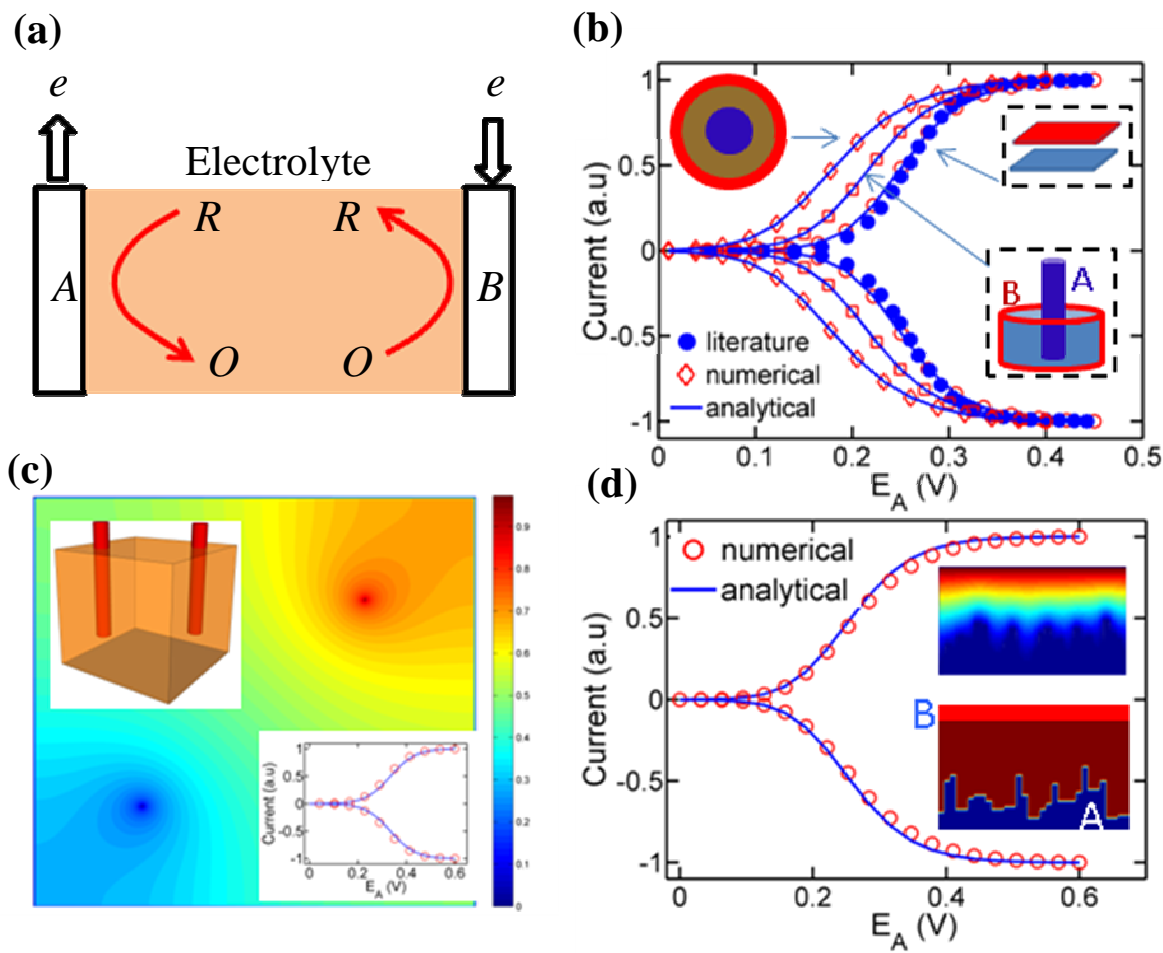

**Figure 6.** Redox detection of molecules using multielectrode schemes. (a) Schematic of detection. The redox species oxidized at A are reduced at B and are again available for oxidation at A, thus amplifying the signal. (b) comparison of results for planar $\left(W = 300\,nm,\ E_0 = 0.24V,\ E_B = 0.1V\right)$, cylindrical ($a_0 = 1\,\mu m,\ W = 1\,\mu m,\ E_0 = 0.20V,\quad E_B = 0V$), and spherical electrodes $\left(a_0 = 0.25\,\mu m,\ W = 0.5\,\mu m,\ E_0 = 0.15V,\ E_B = -0.1V\right)$: experimental results from literature are shown in solid symbols (ref.[5], parameters are same as planar systems), numerical simulation results of this work (open symbols), and analytical results (solid line, eq. (23)). The positive and negative branches denote currents at electrodes A and B, respectively. (c) 2D concentration profile of R molecules in a parallel NW $\left(a_0 = 10\,nm,\ W = 1\,\mu m,\ E_B = -0.2V\right)$ configuration (top inset). Eq. (23) predicts the numerical results (bottom inset). (d) Application of eq. (23) for nanostructured electrodes (bottom inset) with average electrode spacing $< W > = 40\,nm$. Top inset shows the molecule distribution profile. $C_{D,SS}$ for this structure was evaluated numerically.

## III (C). Redox detection of molecules at nanogap electrodes

In the previous section, we described the dynamics of reversible reaction at WE embedded in a semi-infinite media by a robust analytical formulation. Apart from using the formulation for characterization of electrode reaction rates, these electrochemical reactions at such electrode surfaces could have many potentially important applications including, for example, molecule detection. The sensitivity of molecule detection can be considerably improved and the signal can be amplified if additional electrodes are introduced[5] (see Fig. 6). These new electrodes fundamentally change the diffusion geometry of the system. For example, consider a two electrode system (Fig. 6): electrode A (area $A_a$) and electrode B (area $A_b$) where the reactions described by Eq. (1)-(2) occur. The transport of the species R and O between the electrodes is still given by eq. (3). Let the potentials $E_A$ and $E_B$ are applied at corresponding electrodes, respectively (corresponding overpotentials: $\eta_A = E_A - E_0$, $\eta_B = E_B - E_0$). We assume that the potentials are such that oxidation occurs at electrode A while reduction occurs at electrode B. Now the signal at electrode A, due to the oxidation of R, will be amplified by a significant factor as all oxidized molecules will undergo reduction at electrode B and are again available for oxidation at electrode A. This scheme is now successfully used to ultra-sensitive detection of redox species. For example, concentration fluctuations of redox molecules (ferrocenedimethanol[5], Dopamine[6]) in nanogap cavities were reported recently. In addition, similar concepts are used for ultra-sensitive detection of DNA[37, 38]. Once again, however, the analysis of the problem has been limited only for simple electrode configurations[5], while others require complicated numerical solutions[15].

We now extend the analytical formalism developed in Sec. II to address the current amplification in redox detection using nanogap electrodes. We assume that potential sweeps applied, if any, are relatively slow to attain steady state characteristics (i.e., $v \rightarrow 0$). Using eq. (11), the current at both electrodes should be equal at steady state (which implies $d\rho_{s,R}/dt = 0$) and is given as

$$I_{RD} = qC_{D,SS}\left(\rho_{A,R} - \rho_{B,R}\right) = qC_{D,SS}\left(\rho_{B,O} - \rho_{A,O}\right). \tag{21}$$

where the subscript *RD* denote diffusion limited current in a redox detection scheme. The subscripts for the molecule density ($\rho$) denote the electrode location (A or B) and the molecular species (R or O), respectively. This current (eq. (21)) should be equal to the reaction currents given by eq. (5) at the corresponding electrodes, i.e.

$$I_{RD} = A_a q k_0 \left(e^{(1-\alpha)\eta_A}\rho_{A,R} - e^{-\alpha\eta_A}\rho_{A,O}\right) = -A_b q k_0 \left(e^{(1-\alpha)\eta_B}\rho_{B,R} - e^{-\alpha\eta_B}\rho_{B,O}\right). \tag{22}$$

Assuming same diffusion coefficients for R and O, mass conservation indicates that $\rho_R = \rho_{A,R} + \rho_{A,O}$, where $\rho_R$ is the initial concentration of reduced species. Using the eqs. (21)-(22), we obtain

$$I_{RD} = q\rho_R C_{D,SS} \left( \frac{\dfrac{e^{-\alpha\eta_B}}{e^{(1-\alpha)\eta_B} + e^{-\alpha\eta_B}} - \dfrac{e^{-\alpha\eta_A}}{e^{(1-\alpha)\eta_A} + e^{-\alpha\eta_A}}}{1 + \left(\dfrac{C_{D,SS}}{k_0}\right)\left(\dfrac{A_a^{-1}}{e^{(1-\alpha)\eta_A} + e^{-\alpha\eta_A}} + \dfrac{A_b^{-1}}{e^{(1-\alpha)\eta_B} + e^{-\alpha\eta_B}}\right)} \right). \qquad (23)$$

To compare the model (eq. (23)) with results from literature and numerical simulations, we assume that while $E_B$ is held fixed at a potential much lower than $E_0$ (so that O molecules are reduced to R at B), $E_A$ is swept from a low to high bias (much greater than $E_0$). Figure 6 show that our approach predicts the response for a wide variety of electrode configurations. Specifically, Fig. 6b indicates that eq. (23) accurately predict the results for planar nanogap electrodes, as reported in ref. [5]. The same formalism, with appropriate $C_{D,SS}$ given in Table 1, anticipate the results for concentric cylindrical and concentric spherical electrodes (Fig. 6b). Note that as steady state conditions are implied, the positive and negative branches of current in Fig. 6b (i.e., the currents at electrodes A and B, respectively) are of the same magnitude. The application of eq. (23) is not limited to regular electrode configurations shown in Fig. 6b. For instance, it readily predicts the behavior of even complex systems like isolated nanowire (NW) electrodes (with $C_{D,SS} = \pi D \left( \log\left( W/a_0 + \sqrt{\left(W/a_0\right)^2 - 1} \right) \right)^{-1}$, where W is the separation between NW electrodes and $a_0$ is the radius of the NW). The method can be extended to even random nanostructured electrodes, whose $C_{D,SS}$ is not known analytically. With $C_{D,SS}$ a priori numerically estimated, Fig. 6d illustrates that eq. (23) accurately anticipates the numerical simulation results (see Appendix D for details).

The amplification achieved by such multi-electrode schemes for sensitive detection of biomolecules can be understood in simple terms using eq. (23). For $E_A \gg E_0$ and $E_B \ll E_0$, the maximum achievable redox current is

$$I_{RD,max} = q\rho_R C_{D,SS}. \qquad (24)$$

Note that this current is time independent (for closely spaced electrodes) in contrast to the diffusion limited transient current, given by eq. (18), $I(t) = q\rho_R C_{D(t)}$. For planar structures, using $C_{D,SS}$ from Table 1, eq. (24) reduces to $I_{RD,max} = qD\rho_R A_{planar}/W$, ($A_{planar}$ is the area of the electrodes and W the spacing). The second electrode limits the diffusion distance to W and hence converts the transient $t^{-1/2}$ response to a steady state current. Another interesting aspect of the solution is that: Sec. III.A indicates that the current density at a planar electrode is much lower compared to a spherical electrode due to the diffusion limited transport of reactants[3]. Eq. (24) predicts that in a redox scheme, the current density for planar electrodes $(I_{RD.max}/A_{planar} = qD\rho_R/W)$ is exactly the same as the steady state diffusion limited current density at a spherical electrode of similar dimensions (with $a_0=W$, refer Sec. III.A). Hence by using nanogap planar electrodes, one can achieve higher current densities similar to nanoscale spherical electrodes. Equation (24) indicates that any electrode geometry that maximizes $C_{D,SS}$ will also maximize the signal in a redox detection scheme. The ideal candidate in such scenario, given the advantages of massively parallel VLSI fabrication techniques, is interdigitated nanoscale gap electrodes[15]. An interdigitated electrode array can increase $C_{D,SS}$ through two factors: By placing the electrodes closer, W is reduced thus increasing $C_{D,SS}$. Close spacing also allows higher density of electrodes to be placed in a given region which also increases $C_{D,SS}$.

## III (D). Scanning electrochemical microscopy (SECM)

The formalism described above for signal amplification due to redox kinetics in nano electrodes can be extended to other systems as well. A well known example is the SECM[7]. In this scheme, a micro/nano electrode is used to probe an electrochemically active substrate[7, 8] (Fig. 7a). Depending on the applied bias, redox reactions can occur both at the micro/nano electrode and the substrate. It is obvious that the signal characteristics will depend on the geometry of the electrode, the substrate and the spacing between them[7]. Significant research has been devoted to studying SECM experiments, again mostly through numerical analysis with the use of analytical expressions confined to curve fit of numerical data[16, 17]. Our model (eqs. (23)-(24)) provides a general analytic solution for SECM, which can be used to interpret experiments with a wide variety of electrode/substrate combinations.

For any electrode/substrate combination, the only parameter (apart from the reaction coefficients of the SECM tip and substrate) required to predict SECM characteristics (see eqs. (23)-(24)) is the appropriate $C_{D,SS}$, which is a far simpler problem compared to the numerical simulation of eqs. (2), (3), and (6). Moreover, excellent analytical formulations are available in literature for calculating the diffusion equivalent capacitance of various electrode/substrate combinations[24]. This is clearly illustrated in Fig. 7b which compares the prediction of eq. (24) and 3D simulation results from literature[16] for a micro disk SECM. In the vicinity of a conducting substrate, the effective $C_{D,SS}$ of a micro disk electrode is given by $C_{D,SS}(W) \approx 4Da_0 + \pi Da_0^2/W$, where W is the spacing from the conducting substrate[39]. Hence, using eq. (24) (i.e., for $E_A \gg E_0$ and $E_B \ll E_0$), the relative increase in SECM current for a disc electrode is given as

$$\frac{I_{SECM}(W)}{I_{SECM}(W \to \infty)} = \frac{C_{D,SS}(W)}{C_{D,SS}(W \to \infty)} = \frac{\pi a_0}{4W} + 1, \qquad (25)$$

where $I_{SECM}$ denote the current in the presence of a conducting substrate. Figure 7b shows that the analytic results match well with numerical simulation results in literature[16], confirming the generic appeal of the formalism developed in this article for current transients at various microelectrodes.

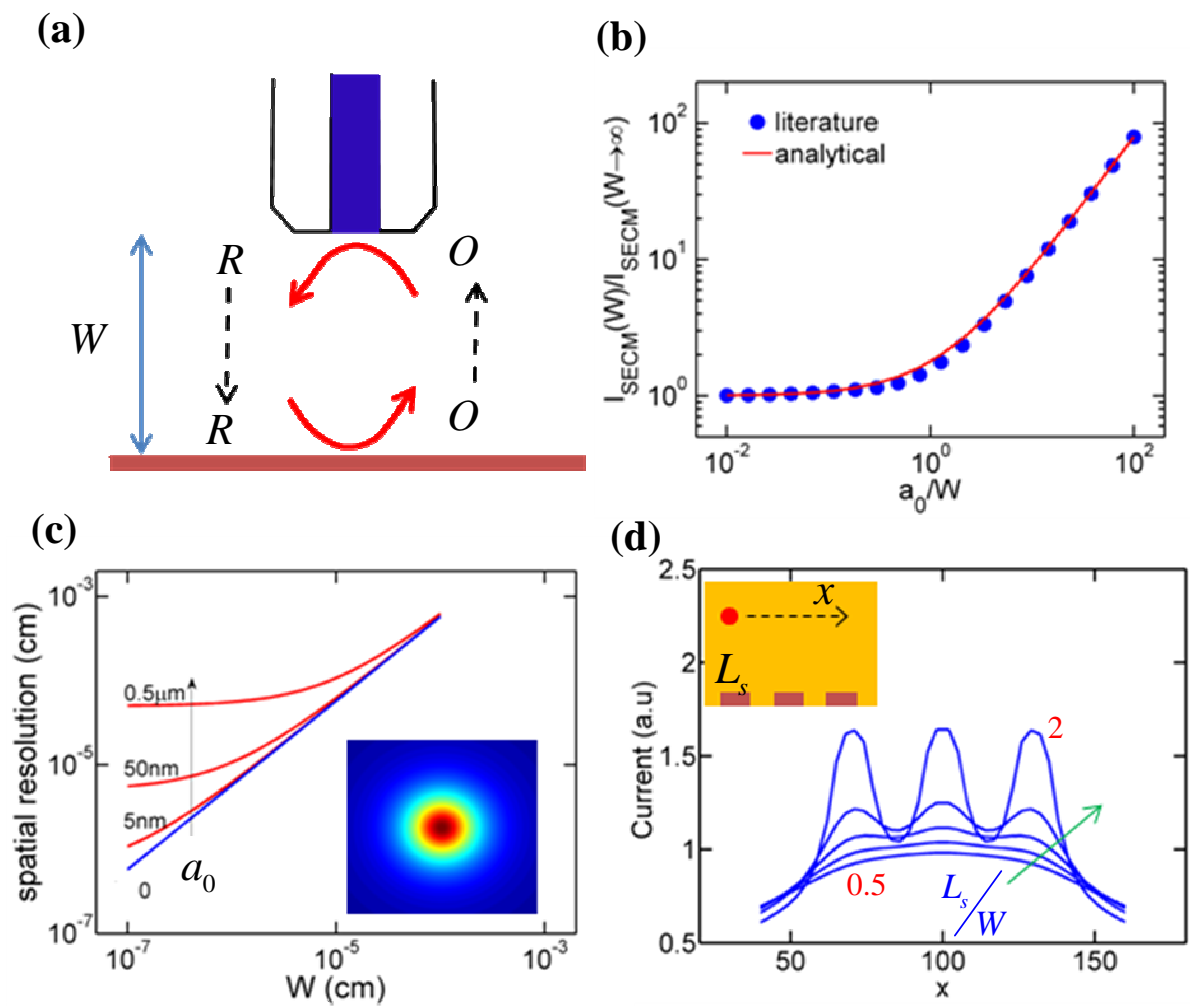

**Figure 7.** Application of the model to SECM. (a) Schematic of SECM. Redox reactions at the SECM tip (bias- $E_A$) and substrate (bias- $E_B$) generate a feedback current to probe the substrate properties. (b) comparison of model (solid line) with results from literature (for $E_A \gg E_0$ and $E_B \ll E_0$, symbols are from ref. [16]) (c) variation of spatial resolution of SECM as a function of probe spacing from substrate. Inset shows the spatial profile of redox current density in substrate. (d) Minimum probe spacing from substrate to distinguish and array of feature size $L_s$ (left inset). As predicted by the model, for $L_s/W=2$, the peaks and the valleys can be clearly resolved.

Nair and Alam, Purdue University (2011)

Further illustrating the relevance of our formalism to SECM, let us consider that problem maximum achievable spatial resolution by SECM. It is intuitively obvious that to maximize the spatial resolution, the SECM probe has to infinitesimally small. For such an ideal probe at a spacing W from the substrate (see Fig. 7a), the relative current density from a point at a distance 'r' along the substrate is given by $J(r) = (W/2\pi)(W^2 + r^2)^{-3/2}$ (obtained through a method of image analysis for $C_{D,SS}$ of an ideal SECM probe above a planar substrate[24]). Defining the spatial resolution, $r_{s,ideal}$, as the region which provides a certain factor $\alpha$ ($0 < \alpha < 1$) of the total current, we have $r_{s,ideal} = W(1-\alpha)^{-1}(2\alpha - \alpha^2)^{0.5}$. For probes with finite size $a_0$, the spatial resolution is $r_{s,finite} = r_{s,ideal} + a_0$. Figure 7c shows the variation of spatial resolution with SECM spacing for various $a_0$. It is clear that even for ideal probes ($a_0 = 0$), the maximum achievable spatial resolution is directly proportional to W. This analysis is particularly useful to determine the spacing required to "image" surface topology using SECM. To probe a surface with feature size $L_S$ (see the inset of Fig. 7d for an illustration involving interdigitated array electrode scheme) with $\alpha = 0.5$, we find that $L_S/W = 2$. As W is changed in different scans from left to right, the current peaks reflect the surface topology and as predicted, at $L_S/W = 2$, the peaks and valleys in current can be distinguished.

## IV. Conclusions

To summarize, we have developed a comprehensive, analytical formalism to understand and predict the behavior of micro and nanostructured electrodes under a wide variety of experimental conditions. The geometry of diffusion and electrodes along with the Butler-Volmer formalism for redox reactions occurring at electrode surfaces is uniquely captured in a closed form, compact analytic expression. Our model indicates that the characteristics of any electrode geometry can be predicted using a single parameter, $C_{D,SS}$, the diffusion equivalent capacitance. Through this approach, we could explain diverse experiments like potentiostatic measurement, linear sweep voltammetry, redox detection of molecules, and SECM. Our methodology and results have interesting implications for design and optimization of electrochemical systems using nanostructured electrodes like fuel cells, electrochemical batteries, dye sensitized solar cells, etc. While more accurate results can always be obtained for specific systems through detailed numerical simulation, this closed form solution methodology has the potential to provide a simple yet powerful analysis that could anticipate key experimental trends and hence could assist in the design and optimization of new electrochemical systems.

## V. Appendices

**A. Electrostatic analogy for diffusion problems – concept of 'diffusion equivalent capacitance':** Equation (3) represents the transient diffusion limited transport of redox species in the system. However, the time independent diffusion flux towards an electrode (dictated by $D\nabla^2 \rho = 0$) can easily be formulated from an analogy with Laplace's equation of electrostatics (dictated by $\varepsilon\nabla^2\Phi = 0$, where $\Phi$ is the potential, see ref. [21] for a detailed discussions). Note that the diffusion current density $-D\nabla\rho$, is analogous to the electric displacement vector $-\varepsilon\nabla\Phi$ in electrostatic systems. Accordingly, the net diffusion flux towards an electrode is the analogue of total electric charge (Q) in electrostatics systems. However, total charge in electrostatic systems is expressed in terms of electrical capacitances as $Q = C_{elec}(\Phi_1 - \Phi_2)$, where is $(\Phi_1 - \Phi_2)$ the net potential difference between two conductors. The analogous net diffusion flux is given by $I_{diff} = C_{D,SS}(\rho_1 - \rho_2)$, where $C_{D,SS}$ is known as the diffusion equivalent capacitance, and $(\rho_1 - \rho_2)$ is the net concentration difference. By analogy with electrostatic systems, note that $C_{D,SS}$ has the same functional form of electrical capacitance except that the dielectric permittivity is replaced with diffusion coefficient. Note that, however, $C_{D,SS}$ is not an electrical capacitance and its units are different.

As the concept of 'diffusion equivalent capacitance' is not widely used in electrochemistry literature, let us illustrate its general appeal of this formalism with a few simple examples. First, consider the steady state diffusion towards a infinitely long cylindrical electrode of radius $a_0$. Let the concentration of R molecules be $\rho_{s,R}$ at electrode surface and $\rho_R$ at a distance W from the electrode surface. Analytical solution of eq. (2) in cylindrical co-ordinates indicate that the total steady state diffusion flux (per unit length) is given by $I = q\dfrac{2\pi D_R}{log\left(\dfrac{W + a_0}{a_0}\right)}(\rho_R - \rho_{s,R})$. Eq. (11)

under steady state conditions (i.e., with $d/dt \to 0$, the second term on the RHS of eq. (11) becomes zero), precisely predicts the same results when appropriate $C_{D,SS}$ is used from Table 1. Similar analysis for spherical electrodes indicate that eq. (11) accurately predicts the current, although the concentration profile for R species vary as $\sim r^{-1}$, where r is the distance from the spherical electrode. More complex systems are explicitly considered in Sec. III. Note that, this approach uses the wealth of information available in electrostatics literature[24] based on many decades of research and there is no need to solve the original mass transport equation. For those configurations whose electrostatic capacitance (and hence the functional form of diffusion equivalent capacitance, $C_{D,SS}$) is unknown, it can be computed for steady state conditions, with ideal Dirichlet boundary conditions. Again, this is a far simpler problem compared to the solution of transient mass transport equation with Butler-Volmer kinetics as boundary conditions.

**B. Derivation of eq. (12):** Equation (11) represents the transient current at a micro electrode with a surface concentration $\rho_{s,R}$ and bulk concentration at a distance W from the electrode. However, it is still not refined enough to use for electrochemical applications. We will now use a perturbation approach: Assume the second term on the right hand side of eq. (11) is negligible. Following eq. (6), equating the particle flux to reaction flux, we obtain

$$qC_{D,SS}\left(\rho_{W,R}-\rho_{s,R}\right)=A_e q\left(k_F\rho_{s,R}-k_R\rho_{s,O}\right). \tag{26}$$

Application of eq. (6) for particle flux conservation leads to

$$D_R\rho_{s,R}+D_O\rho_{s,O}=D_R\rho_R. \tag{27}$$

Using the eqs. (26)-(27) and the boundary conditions $\rho_{W,R}=\rho_R$; $\rho_{W,O}=0$ (see Fig. 1c), we obtain the surface concentration of R as

$$\rho_{s,R}=\rho_R\left(\frac{\dfrac{C_{D,SS}}{A_e}+\dfrac{k_R D_R}{D_O}}{k_F+\dfrac{C_{D,SS}}{A_e}+\dfrac{k_R D_R}{D_O}}\right). \tag{28}$$

Hence the first term in the right hand side of eq. (11) can now be expressed as

$$qC_{D,SS}\left(\rho_{W,R}-\rho_{s,R}\right)=qA_e\rho_R\left(\frac{k_F}{1+\dfrac{A_e}{C_{D,SS}}\left(k_F+\dfrac{k_R D_R}{D_O}\right)}\right). \tag{29}$$

Replacing the first term on the RHS of eq. (11) by eq. (29), we obtain

$$I\left(t\right)=qA_e\rho_R\left(\frac{k_F}{1+\dfrac{A_e}{C_{D,SS}}\left(k_F+\dfrac{D_R}{D_O}k_R\right)}\right)-qA_e\frac{W}{2}\frac{d\rho_{s,R}}{dt}. \tag{30}$$

Equation (30) describes the steady state current due to the redox reaction (Eq.(1)) at the electrode surface in a diffusion limited regime. It, however, assumes that at a constant density contours for R and O at a distance W from the WE, which enable us to conveniently formulate an expression for the total flux towards the WE (see eqs. (26)-(27)). As the reaction proceeds, the use of such a boundary condition is not sufficient and the transient response assumes significance. Based on the perturbation approach in ref.[23], the transient solution can be obtained by realizing that $W \sim \sqrt{2nDt}$ the diffusion distance, where n is the dimensionality of diffusion (Note that there are alternate definitions for diffusion distance, see Sec. 5.2.1 of ref. [3]). So we need to replace $C_{D,SS}$ in eqs. (29) and (30) by the appropriate $C_{D(t)}$ as shown in Table 1. This completes the derivation of eq. (12).

**C. Derivation of eq. (15)**: For a complete description of current transients at microelectrodes, we need to find an analytic expression for the second term on the RHS of eq. (14). Although the time evolution of $\rho_{s,R}$ is predicted by eq. (28) (with $C_{D,SS}$ replaced by appropriate $C_{D(t)}$, see Appendix B), its time derivative is too complex for compact analytic evaluation. We realize that for those systems in which the time derivative is important (particularly LSV experiments with $D_R=D_O$, see section III.B, also supported by numerical simulation results), eq. (28) indicates that

$$\rho_{s,R} \approx \rho_R \left(1 + k_F/k_R\right)^{-1} = \rho_R \left(1 + e^{\beta f(E_A - E_0)}\right)^{-1}, \tag{31}$$

where $0 < \beta \leq 1$. For $\beta = 1$, the above assumption is equivalent to the Nernst limit for the reaction $R \rightleftarrows O + e$. Note that this approximation is made only to make the analytic expressions more tractable, and if necessary, better solutions can be obtained by using the time dependent form of eq. (28). Time derivative of eq. (31), with $E_A = E_i + vt$, leads to eq. (16). Note that such transient effects are significant only for planar systems, as discussed in Sec. III. B.

**D. Numerical simulations**: Equation (3) for the reactants R and O is numerically solved implicitly using finite difference scheme for spatial domain and second order backward differentiation for time integration[40]. At the electrode surfaces, eq. (6) is implemented as the boundary condition. The various simulation parameters used are (unless otherwise stated): $D = 4.7 \times 10^{-6} cm^{-2}s^{-1}$, $k_0 = 0.2 cm\, s^{-1}$, $E_0 = 0.2V$, $\alpha = 0.5$. The accuracy of numerical simulations is evident from the comparison with similar results from literature, as illustrated in Figures 2-7.

## Acknowledgements

The authors acknowledge computational and financial support from Network of Computational Nanotechnology (NCN), National Institute of Health (NIH), and Materials Structures and Devices Center of the Semiconductor Research Center- Focus Center Research Program (FCRP-MSD).